\documentclass[usenatbib,usegraphicx,onecolumn]{mn2e}
\def\mnras{MNRAS}
\def\apj{Ap.J}
\def\apjl{ApJL}
\def\C{{\rm S}}
\def\N{{\rm N}}

\def\dt{\tilde{\eta}_{\rm HI}}
\def\u{{\bf U}}
\def\th{\vec{\theta}}
\def\x{{\bf x}}
\def\HI{{\rm HI}}
\def\H{{\rm H}}

\def\d{\eta_{\HI}}

\def\n{{\bf \hat{n}}}
\def\Tc{T_\gamma}

\def\Ts{T_s}

\def\A10{A_{10}}

\def\rn{r_{\nu}}
\def\rnp{r_{\nu}^{'}}

\def\k{{\bf k}}
\def\kp{k_\parallel}
\def\kpr{{\bf k}_\perp}

\def\m{{\bf m}}

\def\k{{\bf k}}
\def\y{{\bf y}}

\def\u{{\bf U}}
\def\kon{{\bf k_{1 \perp}}}
\def\ktn{{\bf k_{2 \perp}}}
\def\kthn{{\bf k_{3 \perp}}}
\def\kop{{k_{1 \parallel}}}
\def\ktp{{k_{2 \parallel}}}
\def\kthp{{k_{3 \parallel}}}
\def\x{{\bf x}}

\def\HI{{\rm HI}}
\def\H{{\rm H}}

\def\Tc{T_\gamma}
\def\Ts{T_s}
\usepackage{epsfig}
\usepackage{graphics}

\begin{document}
\title[Probing non-Gaussian features in HI at  reionization]
{Probing non-Gaussian features in the HI distribution at the
  epoch of reionization}
\author[S. Bharadwaj and S. K. Pandey]{Somnath
  Bharadwaj$^1$\thanks{Email: 
    somnathb@.iitkgp.ac.in} and Sanjay K.  Pandey$^2$\thanks{Email: 
    spandey@iucaa.ernet.in} 
  \\$^1$ Department of Physics and Meteorology \&
  Centre for Theoretical Studies ,  IIT Kharagpur,  Pin: 721 302 , 
  India \\
$^2$ Deptt. of Mathematics, 
L.B.S.College, Gonda 271001 , India}
\maketitle
\begin{abstract}
The HI distribution  at the epoch of reionization (EOR) is largely
determined by the sizes and distribution of the ionized regions. In
the scenario where the ionized regions have comoving sizes of the
order of a few Mpc, the large scale statistical properties of the HI
distribution are dominated by the Poisson noise of the discrete
ionized regions,  and it is highly non-Gaussian. 
We investigate the possibility of probing reionization by studying these
non-Gaussian features using  future  radio interferometric
observations of redshifted 21 cm HI 
radiation. We develop a formalism relating correlations between the  
visibilities measured at  three different  baselines and frequencies to
the bispectrum of HI fluctuations. For visibilities at the same
frequency, this signal is found to be of the same order 
as the two visibility  correlation  which probes the 
HI power spectrum.  For visibilities at different frequencies, we find 
that the  correlations decay within a frequency difference of $\sim 1
\, {\rm MHz}$. This  implies  that it is, in principle, straightforward to
extract this  HI signal from various contaminants which are believed
to 
have  a continuum spectra and are  expected to be correlated
even at large frequency separations.  

\end{abstract}
\section{Introduction:}
There has recently been a lot of interest in understanding exactly how
and when the universe was reionized. There now are significant
observational constraints mainly from three different kinds of
observations.  The observation of quasars at redshift $ z \sim 6$ which
show strong HI absorption  \citep{becker} indicates that at least  
$1\%$  of the total hydrogen mass   at $ z \sim 6 $  is neutral
\citep{fan}, and the  neutral mass fraction decreases rapidly   at
lower redshifts. This is a strong indication that the 
epoch of reionization ended at $ z\sim 6 $ . Observations of the CMBR
polarization, generated through Thomson 
 scattering of  CMBR photons by free electrons along the line of
 sight, indicates that the reionization began at a redshift $z > 14$.
 On the other hand, the observed anisotropies of the CMBR indicate
 that the total optical  depth of the Thomson scattering is not
 extremely high, suggesting that reionization could not have started at
 redshift much higher than about  30 (\citealt{kogut};
 \citealt{spergel}). 
A third constraint comes from determinations of the IGM temperature
from observations of the ${\rm Ly} \alpha$  forest in the $z$ range
$2$ to $4$ which indicates a complex reionization history with there
possibly being an order unity change in the neutral hydrogen fraction
at $z \le 10$ (\citealt{theuns}; \citealt{hui}).

Mapping the HI distribution at high redshifts using
radio observations  of the redshifted 21 cm radiation
(\citealt{madau};  \citealt{scott};  \citealt{kumar} ) holds the
possibility of probing  the transition from a largely neutral to a
largely ionized universe    at a level of detail surpassing any other
techniques.     \citet{zald} (hereafter ZFH) have developed a
statistical technique based on  the angular power spectrum, on lines
similar to the analysis of CMBR anisotropies, for analysing the HI
signal from the epoch of reionization (EOR) in radio interferometric
observations. Extracting the HI signal from various Galactic and extra 
Galactic contaminants (eg.  \citealt{cooray};  \citealt{dimat};
   \citealt{gnedin2};   
\citealt{oh};  \citealt{dimat1}; \citealt{shaver}; ) is one of the
most important challenges. Most of the known contaminants are expected
to have continuum spectra, and ZFH show that it should in principle be
possible to extract the HI signal using the fact that, unlike the
contaminants,  it   will be 
uncorrelated at two slightly different frequencies. The frequency
dependence of the angular power spectrum of the HI signal and
foregrounds has recently  been  analysed in detail by \citet{santos}.   

An alternative statistical technique for analysing the HI signal is to
study  the correlations between the complex visibilities measured at 
different baselines and frequencies in radio-interferometric
observations. This has been developed in the context of observing HI
from $z < 6$  (\citealt{bharad2}; \citealt{bharad3};
\citealt{bharad4}) and later generalized to the  EOR signal in 
\citealt{bharad6} (hereafter BA). The possibility of using visibility
correlations to quantify the EOR signal  has also been proposed by
\citet{morales}   who further 
discuss how  the different frequency signatures of the contaminants
and the HI signal can be used to distinguish between the two. 
Recently  \citep{morales1} has   addressed the issue of the power
spectrum sensitivity   of the EOR HI signal. 

Various investigations (eg. ZFH, BA) show that the power spectrum of
HI fluctuations at EOR has contributions from  mainly  two
distinct effects, the clustering of the hydrogen which, on large
scales,  is assumed to follow the dark matter distribution and the
fluctuations arising from the presence of discrete  regions of ionized
gas surrounding the sources responsible for reionizing the universe. 
The details of the reionization process  are not very well
understood (eg. \citealt{barkana}), and the shape, size and
distribution of these ionized regions is one of the very important
issues which will be probed by  21 cm HI observations. There has
recently been progress in analytically modeling the growth of the
ionized regions  \citep{furlanetto1} (hereafter FZH)  based on the
findings of simulations   (\citealt{ciardi};  \citealt{sokasiana};  
\citealt{sokasianb}; \citealt{nusser}; \citealt{benson};
\citealt{gnedin}) which show that  there 
will not be a large number of small HII regions around individual
ionizing sources, rather there will be a few large ionized regions  
 centered on places  where the ionized sources are clustered. The size
 of these ionized regions are expected to be  around a few Mpc
 (comoving) or possibly larger at EOR. 
In such a  scenarios, on scales larger than the size of the individual
ionized  regions,  the HI signal will be dominated by the Poisson
noise arising from the discrete nature of the ionized regions
(eg. ZFH, BA, FZH, \citealt{furlanetto2}). Further, the HI signal is 
expected to be highly non-Gaussian .

Nearly all of the work on quantifying    the  EOR HI
signal expected in radio interferometric observations has focused on
the two point statistics namely the angular power spectrum and  the
correlations between pairs of visibilities. Both these quantities are
actually equivalent  and they  basically probe the power spectrum of
HI fluctuations at EOR. The power spectrum completely quantifies a
Gaussian random field, but the higher order statistics would contain
independent information if the HI fluctuations at EOR were not a
Gaussian random field.  FZH  have used
the pixel distribution function, a one-point statistics, to quantify  
non-Gaussian features in the HI distribution. \citet{He} have studied
the non-Gaussian features that arise in the HI distribution in  the
log-normal model. 

 In this paper we address the issue of quantifying the non-Gaussian
 features of  the HI signal expected in radio interferometric
 observations. In particular, we focus on the correlation between
 three visibilities. This is expected to be zero if the signal were a
 Gaussian random field, and deviations from zero are a clear signature
 of the non-Gaussian properties of the HI distribution. Here we 
derive the relation between the three visibility correlation and 
the   bispectrum of the HI  fluctuations. The bispectrum quantifies
correlations between three Fourier modes, and this is non-zero only
 when there are phase correlations between different modes. 
The three visibility correlation, as we show, is comparable to the 
correlations between two visibilities and this leads us to speculate
 that  this  will play an important role in detecting the HI signal.   
Further, the higher order  correlations  contains independent
 information, and observing these would throw independent light
 on  the topology   and morphology of the   HI distribution  at EOR.

Finally, an outline of the paper. In Section 2. we present the
formalism relating the three visibility correlation to the HI
bispectrum.  In Section 3. we introduce a simple model for the HI
distribution at reionization and calculate its bispectrum. 
In Section 4. we present results  for the  three visibility
correlation expected from HI at reionization and discuss some 
consequences. 

It may also be noted that we use the values
$(\Omega_{m0},\Omega_{\lambda0},  \Omega_b h^2,h)=(0.3,0.7,0.02,0.7)$
for the cosmological parameters thoughout. 

\section{Formalism for three visibility correlation}
In this section we follow the notation used in BA  which
also contains a more detailed discussion of the formalism for
calculating the HI signal. The HI radiation at frequency $1420 \, {\rm
  MHz}$ in the rest frame of the hydrogen is redshifted to a 
frequency $\nu=1420/(1+z) \, {\rm MHz}$ for an observer at
present.  The expansion of the universe and the HI peculiar velocity both
contribute to the redshift.   Incorporating these effects, 
the specific intensity $I_{\nu}(\n)$ of redshifted 21 cm HI radiation
 at frequency $\nu$ and direction $\n$ can be written as
 $I_{\nu}(\n)=  \bar{I}_{\nu}(z) \times \d(\n,z)$ 
where 
\begin{equation}
\bar{I}_{\nu}=2.5 \times 10^2 \, \frac{\rm Jy}{\rm sr}  \,
  \left(\frac{\Omega_b   h^2}{0.02}\right)  \left(\frac{0.7}{h}
  \right) \frac{H_0}{H(z)}    
\label{eq:a1} \,.
\end{equation}
and 
\begin{equation}
\d(\n,z)= \frac{\rho_{\HI}}{\bar{\rho}_{\H}}
 \left(1-\frac{\Tc}{\Ts}  \right)  \left[1-\frac{(1+z)}{
 H(z)}\frac{\partial v}{\partial  r}\right]  \ .   
\label{eq:a2}
\end{equation} 
It should be noted that the terms on the right hand side of equations 
(\ref{eq:a1}) and (\ref{eq:a2}) refer to the  epoch when the HI radiation
  originated.  
Here $H(z)$  the Hubble parameter,    ${\bar{\rho}_{\H}}$  the mean 
cosmological  density of hydrogen   and $r$ (or $\rn$)  the
comoving  distance to  the HI calculated ignoring peculiar velocities, 
depend only on $z$.   
The quantities  $\rho_{\HI}$ the  HI density, $\Tc$  the CMBR
temperature, $\Ts$  the HI  spin  temperature and $v$ the radial
component of the HI  peculiar  velocity also vary   with 
position and should be evaluated at   $\x\, = \rn \n$  {\it ie.} the
position where the radiation originated.  It may be noted that
$\d(\x,z)$,  the $21 {\rm cm}$ radiation  efficiency, was  originally
introduced by \citet{madau} who did not include peculiar
velocities. As shown in BA, equation (\ref{eq:a2}) includes  an extra
term  which arises when  the effect of  the  HI  peculiar   velocities
are included. 
The quantity   $\d(\n,z)$ 
incorporate the  details of     the HI  evolution  including effects
of heating, reionization and density fluctuations due to structure
formation.   

 We next introduce $\dt(\k,z)$, the Fourier
transform of  $\eta_{\HI}(\y,z)$, 
\begin{equation}
\d(\y,z)=\int \frac{d^3 k}{(2 \pi)^3} e^{-i \,  \k \cdot \y } 
 \dt(\k,z)  \, . 
\label{eq:a3a}
\end{equation}
 where $\y$ refers to an arbitrary comoving position. Using this we
 can express $\d(\n,z)$ as  
\begin{equation}
\d(\n,z)=\int \frac{d^3 k}{(2 \pi)^3} e^{-i \,  \k \cdot \rn \n} 
 \dt(\k,z)  \,
\label{eq:a3}
\end{equation}
where it is understood that this  refers to the position $\x = \rn \,
\n$. 
  
The ensemble average of various products of $\dt(\k_1,z)$ are used to 
quantify the  statistical properties of the fluctuation in the HI
distribution. We first consider    the  HI power spectrum
$P_{\HI}(\k_1,z)$  defined through   
\begin{equation}
\langle \dt(\k_1,z) \, \dt(\k_2,z) \rangle = (2 \pi)^3 \,  
\delta^3_D(\k_1+\k_2) \, P_{\HI}(\k_1,z)
\label{eq:a4}
\end{equation}
 where $\delta^3_D$ is the three dimensional Dirac delta function.
The power spectrum completely quantifies all properties of the HI
distribution if the fluctuations are a Gaussian random field. The
higher order statistics contain independent information if the
fluctuations are not a Gaussian random field. Here we proceed one step
beyond the power spectrum and also consider 
 the HI  bispectrum $B_{\HI}(\k_1,\k_2,\k_3,z)$    defined through    
\begin{equation}
\langle \dt(\k_1,z) \, \dt(\k_2,z)  \dt(\k_3,z)  \rangle = (2 \pi)^3 \,  
\delta^3_D(\k_1+\k_2+\k_3) \, B_{\HI}(\k_1,\k_2,\k_3,z) \,. 
\label{eq:a5}
\end{equation}

We next mention  a few well known properties of the power
spectrum and bispectrum which are relevant to the  discussion. 
The fact that not all modes are correlated, reflected in the Dirac
delta functions in eq. (\ref{eq:a4}) and (\ref{eq:a5}), is a
consequence of the assumption that HI fluctuations are statistical
homogeneous.  Further, $P_{\HI}(\k)$ is isotropic {\it ie.} does not
depend on the direction of $\k$,  if the effects of the peculiar
velocity are ignored. The redshift space distortion
caused by the peculiar velocities breaks the isotropy of  $P_{\HI}(\k)$
which now depends on the orientation of $\k$ with respect to the  
line of sight. 
Similarly, ignoring redshift space distortions,
$B_{\HI}(\k_1,\k_2,\k_3)$ depends only on the triangle formed by the
wave vectors $\k_1, \k_2$ and $\k_3$, and this  is completely
specified by  the magnitude of the three vectors $(k_1,k_2,k_3)$.  The
bispectrum  also depends on how the triangle is  oriented  with
respect to the line of sight if redshift space distortions are 
included. Finally, we  note that both the power-spectrum and the
bispectrum 
are real quantities. While the power  spectrum is necessarily
positive, there is  no such restriction on  the  
bispectrum.  

We now shift our attention to  radio interferometric observations of
redshifted HI  using an array of low
frequency radio antennas distributed  on a plane. The  antennas all
point in the same direction  ${\bf m}$ which 
we take to be  vertically up wards.  The beam pattern  
$A(\theta)$ quantifies how the  individual antenna, pointing up wards,
responds to signals from different directions in the sky. This is
assumed to be a Gaussian $A(\theta)=e^{-\theta^2/\theta_0^2}$ with 
$\theta_0 \ll 1$ {\it i.e.} 
the beam width of  the  antennas (in radians) is small,
and the part of the sky which contributes to the signal can be  well
approximated by a plane.  In this approximation the unit vector $\n$
can be represented by $\n={\bf m}+\th$, where $\th$ is a two
dimensional vector in the plane of the sky. Using this the angular
fluctuations in the  specific intensity  $\delta I_{\nu}$  can be
expressed as   
\begin{equation}
\delta I_{\nu}(\n)= \bar{I}_{\nu} \int \frac{d^3 k}{(2 \pi)^3} e^{-i
 \,\rn \, (\kp+  \kpr  \cdot \th)}  
 \dt(\k,z)  \, 
\label{eq:a6}
\end{equation} 
where $\kp=\k \cdot {\bf m}$ and $\kpr$ are respectively the
components of $\k$ parallel and perpendicular to  ${\bf m}$. The
component $\kpr$ lies in the plane of the sky. 

The quantity  measured in interferometric  observations is the complex  
visibility  $V(\u,\nu)$ which is recorded for every independent pair
of antennas at every frequency channel in the band of
observations. For any pair of antennas, $\u={\bf d}/\lambda$
quantifies the separation ${\bf d}$ in units of the wavelength
$\lambda$, we  refer to this dimensionless quantity $\u$ as a
baseline. A typical radio interferometric array simultaneously 
measures visibilities at a large number of baselines and frequency
channels, and 

\begin{equation}
V(\u,\nu)= \int d^2 \theta  A(\th) \,  I_{\nu}(\th) \, e^{- i
2 \pi \u \cdot \th} \, .
\label{eq:a7}
\end{equation}

The visibilities record only the angular fluctuations in
$I_{\nu}(\theta)$ and the visibilities arising from  angular
fluctuations  in the  HI radiation are
 \begin{equation}
V(\u,\nu)=\bar{I}_{\nu} \int \frac{d^3 k}{(2 \pi)^3} a(\u -
  \frac{\rn}{2 \pi} \kpr ) \dt(\k,z) e^{-i \kp \rn}
\label{eq:a8}
\end{equation} 

where  $a(U)$ the Fourier transform of the antenna beam pattern
$A(\theta)$, which for a   Gaussian beam
$A(\theta)=e^{-\theta^2/\theta^2_0}$  gives   the Fourier 
transform also to be  a  Gaussian  
$a(\u)=\pi \theta_0^2 \exp{\left[-\pi^2 \theta^2_0 U^2 \right]}$ 
which we use in the rest of this paper. 

In this paper we quantify the statistical properties of the quantity
measured in radio-interferometric observations, namely the
visibilities at different baselines and frequencies. 
Further, we study their relation  to the statistical properties of the
HI distribution.    To this end, we introduce the notation 
\begin{equation}
\C_2(\u_1,\u_2,\Delta \nu)=\langle V(\u_1,\nu + \Delta \nu) V(\u_2,\nu)
\rangle 
\end{equation}

and 
\begin{equation}
\C_3(\u_1,\u_2,\u_3,\Delta \nu_1,\Delta \nu_2)=\langle V(\u_1,\nu +
\Delta \nu_1) V(\u_2,\nu + \Delta \nu_2)  V(\u_3,\nu)  \rangle  
\end{equation}
to denote the correlations between  the visibilities at  different
 baselines  and   frequencies. It should be noted that although we
 have shown $\C_2$ and  $\C_3$ as explicit functions of only the 
frequency differences $\Delta \nu$, all these correlations also depend
 on the the central value $\nu$ which is not shown as an explicit
 argument. Further, throughout our analysis we assume that all
 frequency differences are much   smaller than the central frequency
 {\it ie.} $\Delta \nu/\nu \ll 1$.  

The correlation $\C_2(\u_1,\u_2,\Delta \nu)$ between  the visibilities
 at two    baselines  and   frequencies  has been calculated
 earlier (\citealt{bharad2}; \citealt{bharad3}; BA) who
 find that $\C_2 \sim 0$ if $\u_2 \neq - \u_1$. This is a consequence
 of the statistical  homogeneity of  the HI fluctuations. It is
 sufficient to restrict the analysis to $\u_1=-\u_2 
 =\u$  which we denote as $\C_2(\u,\Delta \nu)$,  and we have 
\begin{eqnarray}
\C_2(U, \Delta \nu) = 
\frac{\bar{I}^2_{\nu} \theta_0^2}{2 \rn^2} 
 \int_0^{\infty}
d \kp \, P_{\HI}(\k)  
\cos(\kp \rnp \Delta \nu) \,.
\label{eq:a9}
\end{eqnarray}
were $\k=k_{\parallel} {\bf m} +  (2 \pi/\rn) \u$ and $\rnp=d \rn/d
\nu$.   The vector $\k$  has  components $k_{\parallel}$ and $(2 
\pi/\rn) \u$ 
respectively parallel  and perpendicular to the line of sight. The 
fact that  $P_{\HI}(\k)$, which includes redshift distortion,   
is isotropic in the directions perpendicular to the line of sight
implies that $\C_2$ is isotropic in $\u$ and we can write
$\C_2(U,\Delta 
\nu)$. We also note that $\C_2$  is real for the HI signal. This follows
from the fact that $P_{\HI}(\k)$ is real and it is unchanged if
$k_{\parallel}  \rightarrow -k_{\parallel}$.

The correlation of the visibilities at three different baselines and
frequencies, $\C_3$   is the quantity of interest in this 
 paper. This will be related to the HI  bispectrum, Here, as for the
 power spectrum, we assume that $\Delta \nu/\nu \ll 1$ ,  whereby  the  only
 term in  eq.  (\ref{eq:a8}) for the  visibility  $V(\u,\nu + \Delta
 \nu)$  which is affected by $\nu \rightarrow \nu + \Delta \nu$ is  
$ e^{-i \kp r_{\nu + \Delta \nu}}$,  which can be approximated as
 $e^{-i  \kp (\rn + \rnp\Delta \nu)}$.  

We then have 
\begin{eqnarray}
&& \C_3(\u_1,\u_2,\u_3,\Delta \nu_1,\Delta \nu_2) 
= \frac{\bar{I}^3_{\nu} }{(2 \pi)^6}
\int d^3  k_1 \ d^3  k_2 \ d^3   k_3 \ a(\u_1-\frac{\rn}{2 \pi} \kon) \,
a(\u_2-\frac{\rn}{2 \pi} \ktn) \,  
\times   \nonumber \\ 
&&  a(\u_3-\frac{\rn}{2 \pi} \kthn) \, 
e^{-i (\kop + \ktp + \kthp) \rn} \, e^{-i (\kop \,  \Delta 
 \nu_1 + \ktp  \,  \Delta \nu_2) \rnp}\, 
 \delta^3_D(\k_1+\k_2 + \k_3) \, 
B_{\HI}(\k_1,\k_2,\k_3)
\label{eq:b1}
\end{eqnarray}
It is convenient to write the $d^3 k$ integrals as $d \kp \, d^2
k_{\perp}$  and integrate over  $d \kthp$,  whereby the 
term $ e^{-i (\kop + \ktp + \kthp) \rn}$ drops out  because of 
the Dirac delta function. Also, we  introduce a new
variable $\y=\k - (2 \pi/\rn) \u$ and use the explicit form for
the function $a(\u)$,  whereby we have 
\begin{eqnarray}
&& \C_3(\u_1,\u_2,\u_3,\Delta \nu_1,\Delta \nu_2) 
= \frac{\bar{I}^3_{\nu}}{(2 \pi)^6}  \int d \kop  \ d \ktp
 \   e^{-i (\kop \,  \Delta  \nu_1 + \ktp  \,  \Delta \nu_2)
   \rnp}   \int d^2 y_1 \, \, d^2 y_2 \, \, d^2 y_3  \ \times \nonumber \\  
&&  \delta^2_D[(2 \pi/\rn)  
 (\u_1+\u_2+\u_3)+ \y_1 +\y_2+\y_3] 
(\pi \theta_0^2)^3 \exp[- (\rn \theta_0/2)^2 (y_1^2 + y_2^2 +
 y_3^2)]  B_{\HI}
\label{eq:b2}
\end{eqnarray}
where the arguments of the bispectrum change as we carry out the
integrals,  but we do not show them explicitly.  

Carrying out the $d^2 y_3$ integral we have 
\begin{eqnarray}
&& \C_3(\u_1,\u_2,\u_3,\Delta \nu_1,\Delta \nu_2) 
= \frac{\bar{I}^3_{\nu}}{(2 \pi)^6}  \int d \kop  \ d \ktp
 \   e^{-i (\kop \,  \Delta  \nu_1 + \ktp  \,  \Delta \nu_2)
   \rnp}   \int d^2 y_1 \, \, d^2 y_2  \ \times
 \nonumber \\   
&&  \exp[- (\rn \theta_0/2)^2 (y_1^2 + y_2^2 )] \,  \exp[- (\rn
   \theta_0/2)^2   
\{\y_1+\y_2+(2\pi/\rn)(\u_1+\u_2+\u_3)\}^2] \,  B_{\HI}
\label{eq:b3}
\end{eqnarray}
The point to note is the two Gaussian functions $\exp[- (\rn
 \theta_0/2)^2 (y_1^2 + y_2^2 )]$ and $\exp[- (\rn  \theta_0/2)^2   
\{\y_1+\y_2+(2\pi/\rn)(\u_1+\u_2+\u_3)\}^2]$   are  peaked around
 different values of $\y_1$ and $\y_2$. While the former is peaked
 around $\y_1=\y_2=0$, the latter is peaked around $\y_1+\y_2=
(-2\pi/\rn)(\u_1+\u_2+\u_3)$. The peaks of the two functions have very
 little  overlap if $\mid \u_1+\u_2+\u_3 \mid > 0$, and the visibility
 correlations are exponentially suppressed if the vector sum of the
 baselines differs from zero. There  are  substantial correlations
 only for the sets of baselines    for which   $\mid    \u_1+\u_2+\u_3
 \mid \le  (\pi \theta_0)^{-1}$. In the rest of our analysis we only
 consider combinations of baselines for which   $\u_1+\u_2+\u_3=0$, 
and  the product of the two  Gaussian functions becomes $\exp[- 2 \,
 (\rn  \theta_0/2)^2  (y_1^2 + y_2^2 + \y_1 \cdot \y_2) ]$. This can
 be further simplified if the baselines we are dealing with are much
 larger than $1/(\pi \theta_0)$. We can then approximate  this function by
 a  product of two Dirac delta functions $\approx (16 \pi/3) (\rn 
 \theta_0)^{-4} \delta^2_D(\y_1+\y_2/2)    \delta^2_D(\y_2)$. Using
 this in eq. (\ref{eq:b3}) we have 
\begin{equation}
 \C_3(\u_1,\u_2,\u_3,\Delta \nu_1,\Delta \nu_2) 
= \frac{\bar{I}^3_{\nu} \, \theta_0^2}{12 \, \pi \, \rn^4} 
\int d  \kop \ d  \ktp \, e^{-i (\kop \,  \Delta  \nu_1 + \ktp  \,
 \Delta \nu_2) \rnp} \  B_{\HI}(\k_1,\k_2,\k_3)
\label{eq:s1}
\end{equation}
where 
$\k_1=\kop \m + (2 \pi/\rn) \u_1$, $\k_2=\ktp \m + (2 \pi/\rn) \u_2$ 
and $\k_3=-(\kop + \ktp) \m + (2 \pi/\rn) \u_3$. Further, it can be
verified that $\C_3$ is real,  and 
\begin{equation}
 \C_3(U_1,U_2,U_3,\Delta \nu_1,\Delta \nu_2) 
= \frac{\bar{I}^3_{\nu} \, \theta_0^2}{12 \, \pi \, \rn^4} 
\int  d  \kop \  d  \ktp \,
\cos[(\kop \,  \Delta  \nu_1 + \ktp  \, 
 \Delta \nu_2) \rnp] \  B_{\HI}(\k_1,\k_2,\k_3) 
\label{eq:s2}
\end{equation}
where we have also incorporated the fact that $\C_3$ depends only the
triangle formed by $\u_1,\u_2$ and $\u_3$ which is  completely
specified by just the magnitudes $(U_1,U_2,U_3)$. 

We use eqs. (\ref{eq:a9}) and (\ref{eq:s2}) to calculate the
visibility correlations expected during the epoch of reionization. 

\section{A Model for the HI distribution}
The reionization of the HI in the universe started,  possibly at a
redshift $z \sim 30$, when the first luminous objects were formed.
The radiation from these luminous objects and from the subsequently
formed luminous objects   ionized  the low density HI in the universe.   
The reionization commences in  small spherical regions (Stromgren
sphere) surrounding the luminous objects. These spheres are filled with
ionized HII gas,  the rest of the universe  being filled with HI .
Gradually  these  ionized regions  grow until they finally 
overlap, filling up the whole of space, and all the low density gas in
the universe is  ionized. The HI distribution during reionization is
largely determined by the ionized regions. This  is expected to be
highly non-Gaussian carrying signatures of the size, shape and
distribution of the  discrete ionized regions.   Here we adopt a
simple model for the  ionized regions.  Though simple, this model  
suffices to illustrate the non-Gaussian nature of the HI distribution
and allows us to calculate some of  the salient observable
consequences.  

We assume that the HI gas is heated well before it is reionized, and
that the spin temperature is coupled to the gas temperature with 
$\Ts \gg \Tc$ so that  $(1-\Tc/\Ts) \rightarrow 1$. It then follows
that $\d >0$  (eq. \ref{eq:a2}) {\it ie.} the HI will be seen in
emission. At any epoch a  fraction of the volume $f_V$ is completely 
ionized, the ionized gas being in  non-overlapping spheres of comoving
radius $R$,  the centers of the  spheres being randomly distributed. 
This model is  similar to that used by ZFH  in the context of
HI emission,  and \citet{gruz} and \citet{knox} in the context of the
effect of patchy reionization on the CMBR. 
One would expect the centers of the ionized spheres to be clustered, 
given the fact that we identify them with the locations of the first
luminous objects which are believed to have formed at the peaks of the
density fluctuations.  This effect, included in BA, has
not been taken into account here. 

 Following ZFH,  we   assume that  
 the   mean   neutral fraction  $\bar{x}_{\HI}$ at any epoch is given
 by  
\begin{equation}
\bar{x}_{\HI}(z)=\frac{1}{1+\exp((z-z_0)/\Delta z)}
\end{equation}
with $z_0=10$ and $\Delta z=0.5$ so that $50 \%$ os the hydrogen is
neutral  at a redshift $z=10$. The mean comoving number density of
ionized spheres $\bar{n}_{\HI}$ is related to the quantities defined
earlier as $f_V=1-\bar{x}_{\HI}=(4 \pi R^3/3) \bar{n}_{\HI}$. We have
kept $R$ as a free parameter and have used this to determine
$\bar{n}_{\HI}$.  

We assume that the total hydrogen density traces the dark matter and
hence it is $\bar{\rho}_{\H} (1  + \delta)$  where $\delta$ 
refers to the fluctuations in the  dark 
matter distribution. Then,  in our model, the  
HI density is  $\rho_{\HI}(\x,z)=\bar{\rho}_{\H} (1
+ \delta) \left[ 1 - \sum_a \theta(\mid \x -\x_a \mid/R) \right] $
, where $a$ refers to the
different ionized spheres with centers at $\x_a$, and $\theta(y)$ is the
Heaviside step function defined such that $\theta(y)=1$ for $0 \le y
\le 1$ and zero otherwise. We then have 
\begin{equation}
\d(\x,z)=\left[ 1+\delta-\frac{1+z}{H(z)} \frac{\partial v}{\partial r}
  \right]  \left[ 1 - \sum_a \theta(\frac{\mid \x -\x_a \mid}{R})
  \right] \,. 
\label{eq:c1}
\end{equation}
where $v$ refers to the peculiar velocity caused by $\delta$. 
The point to note is that $\d(\x,z)$ has contributions from two
distinct effects namely the fluctuations arising from the
gravitational clustering of the hydrogen which follows  the dark
matter distribution and the  discrete ionized regions. Earlier studies 
(ZFH) have shown that the contribution from the
discrete ionized regions dominates the HI power spectrum on
length-scales larger than the size of the individual ionized bubbles
at redshifts $z \sim 10$ when $f_V \sim 0.5$  and the HI signal is
expected to be maximum.  In the standard scenario, the initial
dark matter fluctuation  $\delta$ is e assumed to be a Gaussian 
random field for which the bispectrum is zero. Non-Gaussian 
features of order $\sim \delta^2$  arise from   non-linear effects as
the density fluctuation grow,  but these effects are expected to   be
very small on  the length  scales of our interest at redshifts $z \ge
10$. The bispectrum  $B_{\HI}$ too  will be dominated by the
non-Gaussian features  arising from the discrete ionized regions.
Further, we expect the gravitational clustering of the hydrogen to
make a smaller contribution to the bispectrum than it does to the
power spectrum. The aim here being to investigate  the non-Gaussian
effects through a study of the bispectrum, it is justified to focus on
just the contribution arising from the individual ionized regions,
ignoring the effects of gravitation clustering.    Under this
assumption  
\begin{equation}
\d(\x,z)=\left[ 1 - \sum_a \theta(\frac{\mid \x -\x_a \mid}{R}) 
  \right] \,. 
\label{eq:c2}
\end{equation}
and it Fourier transform for $k >0$ is 
\begin{equation}
\dt(\k,z)=\frac{-f_V \, W(k R)}{ \bar{n}_{\HI}} \sum_a e^{i \k \cdot
  \x_a}
\label{eq:c3}
\end{equation}
where $W(y)=(3/y^3)[\sin(y) - y \cos(y)]$ is the spherical
top hat window function. Using these we have 
\begin{equation}
P_{\HI}(\k)=\frac{f_V^2 W^2(k R)}{\bar{n}_{\HI}}
\label{eq:c4}
\end{equation}
and 
\begin{equation}
B_{\HI}(\k_1,\k_2,\k_3)=-\frac{f_V^3 \ W(k_1 R) \ W(k_2 R) \  W(k_3 R)
}{\bar{n}^2_{\HI}} 
\label{eq:c5}
\end{equation}
respectively for the power spectrum and the bispectrum. We use these
to calculate the visibility correlations expected in this model.  

Our model has a limitation that it cannot be used  when a large
fraction of the volume is ionized as the ionized spheres start to
overlap and the HI density becomes negative in the overlapping
regions. Calculating  the fraction of the total volume where the HI
density is negative, we find this to be $ f_V^2/2$. We use this to
asses the range of validity of our   model. We restrict the model to
$z>10$ where $f_V<0.5$, and the HI density is negative in less than
$12.5 \%$ of the total volume.  
\section{Results and Discussion}
In this section we present results for the visibility correlations
expected from HI during  the epoch of reionization. Our aim being
to illustrate the non-Gaussian nature of the expected signal and its
dependence on the   ionized regions, we show results centered on  
  only at a single frequency namely $125 \, {\rm
   MHz}$.  This corresponds to a  redshift $z=10.4$ when  the mean
 neutral fraction is   $\bar{x}_{\HI}=0.67$  ({\it ie.} $f_V
 =0.33$). We choose this particular frequency as $\bar{x}_{\HI}$
 is quite  close to $0.5$ where the HI signal is expected to be
 maximum, simultaneously  ensuring that the volume fraction where the
 HI density  predicted by our model becomes  negative is small ($\sim
 5 \%$).   
Further, the HI signal is expected to be dominated by discrete ionized
regions  and hence we  anticipate significant non-Gaussian features. 

We have used eqs. (\ref{eq:a9}) and (\ref{eq:s2}) to calculate the
expected correlations between two and three visibilities
respectively. For this it is necessary to specify   a value for
$\theta_0$, the   beam size of the   individual antennas in the array.
Further, it may be noted that $\theta_0  \approx 0.6 \times
\theta_{\rm   FWHM}$. The value of $\theta_0$ will depend on the  
physical dimensions of the antennas and the wavelength of
observation.  For the GMRT $\theta_0=1^{\circ}$ at $325 \, {\rm
  MHz}$. We  scale this  using $\theta_0 \propto  \lambda$ to  obtain 
$\theta_0=2.6^{\circ}$    at $125 \, {\rm MHz}$  which we use
here. The HI signal predicted here is  for observations using
the GMRT,  and they  can be directly compared to  those in
BA.     Both $\C_2$ and $\C_3$  scale as
$\theta_0^2$, and it is straightforward to scale the results presented 
here to make  visibility correlation predictions for  other radio
telescopes.  

The comoving radius   of the ionized spheres  $R$  is a free parameter
in  our model. Investigations on the growth of the ionized spheres
(FZH) show that these will be at least a few Mpc in
radius (possibly larger) at the redshift of interest. 
 We have considered three possible values $R=(1, \,  3 ,
\, 5) \, {h^{-1}   \, Mpc}$ for which  the respective 
values of $\bar{n}_{\HI}$ are $(78, \, 2.9 , \, 0.63 ) \, \times
10^{-3} \, h^{3} {\rm Mpc}^{-3}$.   

For ease of graphical presentation, we have restricted our analysis of
$\C_3$ to equilateral triangles for which the size of the baseline $U$
completely specifies the triangle, and we have $\C_3(U,\Delta
\nu)$. Further, we first consider the correlations at the same
frequency {\it ie. } $\Delta \nu=0$   Figures (\ref{fig:1}) and
({\ref{fig:2}) show the results  $[\C_2(U)]^{1/2}$ and 
  $[-\C_3(U)]^{1/3}$ respectively.  

\begin{figure}
\includegraphics[width=84mm]{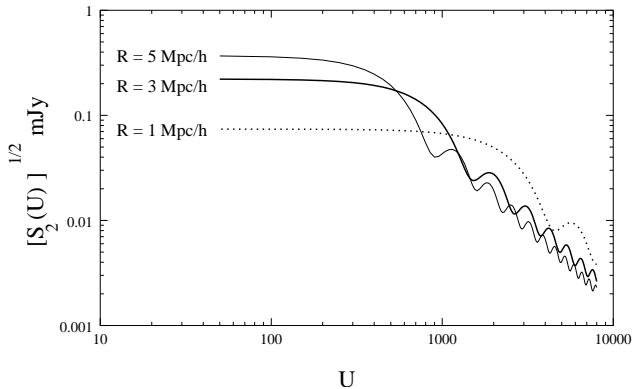}
\caption{This shows the visibility correlation $\C_2(U,\Delta
  \nu)^{1/2}$ as   a function of $U$ for $\Delta \nu=0$, for different
  values of $R$, the comoving radius of the ionized  spheres. These
  predictions are for observations centered at $125 \, {\rm MHz}$. 
}
\label{fig:1}
\end{figure}
\begin{figure}
\includegraphics[width=84mm]{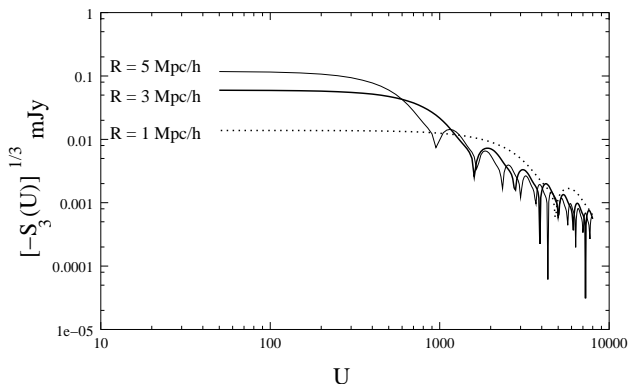}
\caption{This shows the visibility correlation $[-\C_3(U,\Delta
  \nu)]^{1/3}$ as   a function of $U$ for $\Delta \nu=0$, for different
  values of $R$, the comoving radius of the ionized  spheres. These
  predictions are for observations centered at $125 \, {\rm MHz}$. 
}
\label{fig:2}
\end{figure}
We find that at small $U$,   $[\C_2(U)]^{1/2}$ is  more or less
constant with a value  of the   order of $\sim 0.2 \  {\rm
  mJy}$ for $R=3 \, h^{-1}\, {\rm Mpc}$. The   signal is proportional
to $R^{3/2}$ and its 
magnitude  increases as the ionized spheres become larger.  Each
baseline $U$ can be associated with a comoving length-scale  $\rn/(2
\pi U)$
at the comoving distance  where the HI radiation originated.   The
signal from 
the ionized spheres  is constant across the   baselines for which
$\rn/(2 \pi U)$
is larger than the size of the  spheres, and the signal falls at
baselines for which $U > \rn/(2 \pi R)$.  Each    baselines resolves out
features larger than $\rn/(2 \pi U)$, and the  presence of discrete
ionized 
regions make very little contribution to  the signal at the large 
baselines.    Comparing the results for $[\C_2(U)]^{1/2}$ presented
here with those  presented in BA  which also includes the
effects of gravitational clustering, we note that the gravitational
clustering signal is also of the order $\sim 0.1 \,  {\rm    mJy}$ at
small $U$.  The gravitational clustering signal  also falls with
increasing $U$,  and the combined signal would depend critically on
the size of the bubbles.  For example, the signal from discrete
ionized sources would dominate over the gravitational clustering
signal at baselines $U > 500$ if the  ionized spheres had  comoving
radius $R= 5 \, h^{-1} \, {\rm Mpc}$,  whereas the gravitational
clustering signal would possibly dominate throughout 
for  $R= 1 \, h^{-1} \, {\rm Mpc}$.

Turning our attention next to $\C_3$ (Figure \ref{fig:2}), the first
point to note is that this is negative. The shape of $\C_3$ as a
function of $U$ is very similar to that of $\C_2$, and its magnitude
is  around  $[-\C_3(U)]^{1/3} \, \sim \, 0.06 \, {\rm mJy}$ at $R=3 \,
h^{-1} \, {\rm Mpc}$, 
which is around three times smaller than $[\C_2(U)]^{1/2}$. At small
$U$,   $[-\C_3(U)]^{1/3}$ is  more or  
less  constant. Although our results are restricted to equilateral
triangles, we expect the correlations to be nearly constant for
triangles of all shapes provided all the baselines satisfy $U < \rn/(2
\pi R)$.  The   signal is proportional to $R^2$ and 
its magnitude  increases a little faster than that of
$[\C_2(U)]^{1/2}$  as $R$ is increased. We expect the dark matter
density fluctuations at $z > 10$  to be well in the  linear regime  on
comoving  length-scales $\sim 10 \, h^{-1}\,  {\rm Mpc}$ or larger,      
and the contribution to $\C_3$ from  non-linear
gravitational  clustering  is expected to be very small on   
these scales.   It may be noted that the contribution to $\C_3$  
from linear gravitational clustering is exactly zero in the standard
scenario where the initial density fluctuations are a Gaussian random
field.   
Further, the  comoving length-scale      $10 \, h^{-1}\,  {\rm   Mpc}$
corresponds to the baseline  $U \sim 100$ and we expect the
contribution from individual ionized spheres 
considered here to be the dominant signal at these baselines. 
\begin{figure}
\includegraphics[width=84mm]{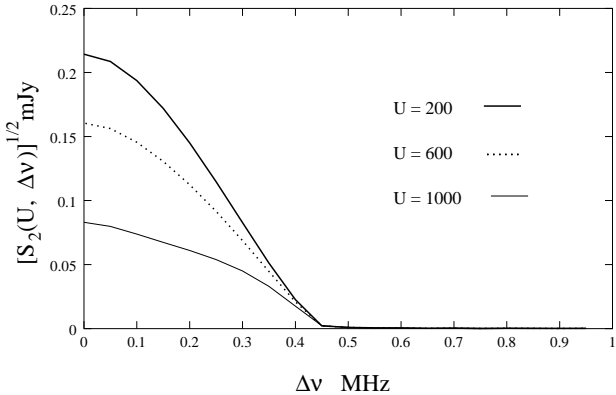}
\caption{This shows the visibility correlation $\C_2(U,\Delta
  \nu)^{1/2}$ as   a function of $\Delta \nu$ for the three different 
  values of  $U$ shown in the figure.  The  comoving radius of the
  ionized  spheres is assumed to be $R=3 \, h^{-1} {\rm Mpc}$ and the
  predictions are for observations centered at $125 \, {\rm MHz}$. 
}
\label{fig:3}
\end{figure}
\begin{figure}
\includegraphics[width=84mm]{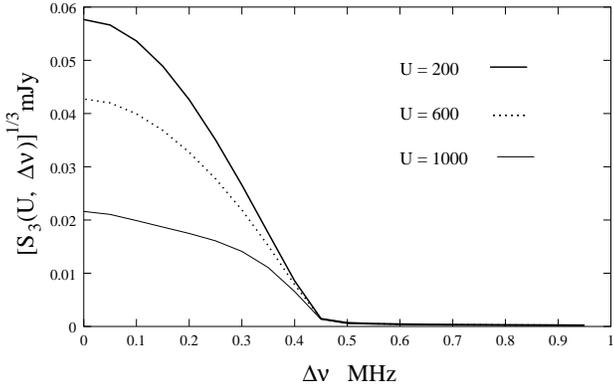}
\caption{This shows the visibility correlation $[-\C_3(U,\Delta
  \nu)]^{1/3}$ as   a function of $\Delta \nu$ for the three different 
  values of  $U$ shown in the figure.  The  comoving radius of the
  ionized  spheres is assumed to be $R=3 \, h^{-1} {\rm Mpc}$ and the
  predictions are for observations centered at $125 \, {\rm MHz}$. 
}
\label{fig:4}
\end{figure}

We next consider the correlations between the visibilities at
different frequencies. Again, for ease of graphical presentation we
have restricted our analysis of $\C_3(U,\Delta \nu_1, \Delta \nu_2)$
to equilateral triangles with the added restriction that $\Delta \nu_1
= \Delta \nu_2 = \Delta \nu$, so we have $\C_3(U,\Delta \nu)$. We have
shown results only for $R=3 \, h^{-1} \, {\rm Mpc}$, but a similar
behaviour is expected for other values also. We find that both
$\C_2(U,\Delta \nu)$ and $\C_3(U,\Delta \nu)$ fall rapidly, in nearly
the same fashion independent of $U$, and are very close to zero by
$\Delta \nu \approx  0.5 \, {\rm MHz}$. 

One of the main challenges in observing cosmological HI is to  extract
it  from  various contaminants which are expected to swamp this
signal. The contaminants include  Galactic synchrotron  
emission, free-free  emission from ionizing halos  \citep{oh}, faint  
radio loud quasars \citep{dimat1} and synchrotron emission from low
redshift galaxy clusters \citep{dimat}. Fortunately, all of these
foregrounds have smooth continuum spectra and we expect their
contribution to the visibilities to be correlated over large $\Delta
\nu$, whereas the HI contribution is  uncorrelated beyond   $1 \, 
   {\rm MHz}$ or less. 
It is, in principle, straightforward to fit the
visibility correlations $\C_2$ and $\C_3$  at large $\Delta \nu$ and
remove any slowly  varying component thereby separating the
contaminants from the HI signal. We also use this opportunity to note
that  
this is a major advantage of using visibility correlations as compared
to the  angular power spectrum  which exhibits substantial
correlations      
even at two frequencies separated by $\sim 10 \, {\rm MHz}$
\citep{santos}.  

An important fact which emerges from our analysis is that the HI
signal  in the correlation between three visibilities is
of the same order as  the correlation between two  visibilities, the
former being around three times smaller. This is a generic feature of
the EOR HI signal, valid   if the ionized regions are bubbles of   
the size $R=1 h^{-1} \, {\rm Mpc}$ or larger. This signal arises from
the Poisson noise of the discrete ionized regions, and it is  enhanced
if the  size of the bubbles is increased. The fact that there is a
substantial $\C_3$ tells us  that there  are large  phase
correlations between the visibilities. This is a consequence of  the  fact
that there are only a few coherent features (the ionized regions)
which dominate the whole HI signal.

Investigations on the growth of ionized regions (FZH)
show that there will be a spread in the  sizes of  the ionized regions
at any given epoch. This will smoothen some of the sharp features seen 
in Figures (\ref{fig:1}) and (\ref{fig:2}). The ringing seen in the
these figures is an artifact of there being only a single value of $R$
and we do not expect this feature to be there if we have a spread in
$R$. Further, the gravitational clustering signal not shown here may
also dominate at large $U$. Despite all these limitations, we can
still expect substantial correlations between three visibilities in a 
a more realistic analysis,  this  
being a robust signature  of the fact that reionization occurs through
a few, large ($R \sim $ a few Mpc) bubbles of ionized gas and the HI 
signal is dominated by  Poisson noise.

We next briefly discuss  the noise levels and the integration times 
required to observe the HI signal, particularly addressing  the
question  whether $\C_3$ can be detected with integration times
comparable to those needed for  $\C_2$. 
We consider an array of $N$ antennas, the observations lasting a time
duration $t$, with frequency channels of  
width $\delta \nu$ spanning a total bandwidth $B$.  It should be noted
that the effect of a finite channel width $\delta \nu$ has not been
included in our calculation which assumes infinite frequency
resolution. This effect can be easily included by convolving our
results for the visibility correlation with the frequency
response function of a single channel. Preferably, $\delta \nu$
should be much smaller than the frequency separation at which the
visibility correlation become  uncorrelated.  We use $S$ to denote the  
frequency separation within which the visibilities are correlated,  and
beyond which they become uncorrelated.

 We use $\N_2$
and $\N_3$ to denote the rms. noise in $\C_2$ and $\C_3$
respectively. It is well known that  $\N_2 = \left( \frac{2 k_B
  T_{SYS}}{A_{ef}}   \right)^2 \frac{1}{\delta \nu \, t}$
\citep{thomp}, and we have $\N_3\sim \left( \frac{2 k_B 
  T_{SYS}}{A_{ef}}   \right)^{3} \frac{1}{(\delta \nu \, t)^{3/2}}$
assuming that we have Gaussian random noise,  
where $T_{SYS}$ is the system temperature and $A_{ef}$ is the
effective area of a single antenna. The noise contributions  will be
reduced by a factor $1/\sqrt{N_o}$ if we  combine $N_o$ independent
samples of the  visibility correlation.  A possible observational
strategy for  a preliminary detection of the HI signal would be to
combine the visibility correlations at all baselines and frequency
separations where there is a reasonable amount of signal. This gives
$N_o=[N(N-1)/2] \,  (B/\delta \nu) \, (S/\delta \nu)$ for the two
visibility correlation and $N_o=[N(N-1)(N-2)/6] \,  (B/\delta \nu) \,
(S/\delta \nu)^2$ for the three visibility correlations.  It should be
noted that we have used the fact that the $\C_3$ is  non-zero only
for the baselines between triplets of antennas. Combining all of this
we have $[\N_2]^{1/2} \sim  \left( \frac{2 k_B
  T_{SYS}}{A_{ef}}   \right) \left[ \frac{2}{N (N-1) B S}
  \right]^{1/4} \frac{1}{t^{1/2}}$ and $[\N_3]^{1/3} \sim  \left( 
\frac{2 k_B   T_{SYS}}{A_{ef}}   \right) \left[ \frac{6}{N (N-1)(N-2)
    B S^2 } \right]^{1/6} \frac{1}{t^{1/2}}$.  The ratio 
$[\N_3]^{1/3}/[\N_2]^{1/2} \sim [N (N-1) B/(N-2)^2 S]^{1/12}$ has a
very weak dependence on $N$, $B$ and $S$ for a reasonable choice of 
values, and is of order unity.  We thus see that, for a given 
integration time,  we will achieve comparable noise levels in both the
two and three visibility correlations.  Estimates of the integration
time to detect $\C_2$ (or equivalently the angular power spectrum)
(BA, ZFH) indicate this to be around a few hundred  hours. We find
that it should be possible to also detect $\C_3$ in a comparable
integration time.  

\section*{Acknowledgments}
SB would like to thank  Sk. Saiyad Ali and Jayaram  Chengalur for
 useful discussions.   SB would also like to acknowledge BRNS, DAE,
 Govt. of India,for  financial support through sanction
 No. 2002/37/25/BRNS.    SKP would like to acknowledge the Associateship
 Program, IUCAA for   support.

\end{document}